# 133-Tbps 1040-km (13×80 km) Lumped-Amplified Transmission Over 22 THz in S-to-U-Band Using Hybrid Multiband Repeater with PPLN-Based Optical Parametric Amplifiers and EDFAs

Shimpei Shimizu[(1)], Takayuki Kobayashi[(1)], Masashi Abe[(2)], Takushi Kazama[(1,2)], Akira Kawai[(1)], Kosuke Kimura[(1)], Masanori Nakamura[(1)], Fukutaro Hamaoka[(1)], Koji Enbutsu[(2)], Takahiro Kashiwazaki[(2)], Munehiko Nagatani[(1,2)], Hitoshi Wakita[(2)], Yuta Shiratori[(2)], Hiroshi Yamazaki[(1,2)], Hiroyuki Takahashi[(1,2)], Takeshi Umeki[(1,2)], and Yutaka Miyamoto[(1)]

[(1)] NTT Network Innovation Laboratories, NTT Corporation, Yokosuka, Japan, shimpei.shimizu@ntt.com
[(2)] NTT Device Technology Laboratories, NTT Corporation, Atsugi, Japan

**Abstract** *We demonstrated 22.05-THz four-band long-haul transmission with a S-to-U-band lumped repeater consisting of PPLN-based optical parametric amplifiers and EDFAs over an 80-km-span SMF link. The achieved net bitrate was 133.06 Tbps at 1040 km with the 25.5-dBm fibre launch power designed by accounting for ISRS.* ©2024 The Author(s)

## Introduction

Wavelength-division multiplexing (WDM) transmission technologies beyond C+L-band have been an area of growing research interest. In particular, the bandwidth extension to shorter wavelength bands such as S- and E-bands has been reported with doped-fibre amplifiers [1–4]. In ≥80-km-span inline-amplified systems oriented to terrestrial fibre networks, however, the transmission distance is limited by the transmission quality of shorter-wavelength channels degraded by the performance of the inline amplifiers and large attenuation due to inter-channel stimulated Raman scattering (ISRS). A distributed Raman amplifier (DRA) is a promising approach to improve the transmission performance in the shorter wavelength bands. Previously, 110.7-Tbps 1040-km (13×80 km) S+C+L-bands transmission with 18.30-THz signal bandwidth using the DRA was reported [1]. In addition, four-bands (E+S+C+L) 26.25-THz transmission was demonstrated over 2×100 km [4]. Meanwhile, as a demonstration of the extension to the longer wavelength band, a C+L+U-bands transmission was reported where long-haul three-bands inline-amplified transmission without the DRA was achieved by effectively utilizing the ISRS effect between the C+L-band and U-band [5]. However, there are no reports of 1000-km-class transmission over >20-THz signal bandwidth beyond the three bands.

In this paper, we demonstrate the first four-band inline-amplified transmission over 1000 km using an S+C+L+U-band with 80-km G.652.D single-mode fibre (SMF) spans. The S-to-U-band repeater consisted of erbium-doped fibre amplifiers (EDFAs) and optical parametric amplifiers (OPAs) based on periodically poled LiNbO$_3$ (PPLN) waveguides. OPAs are another promising solution for amplifying the multiband signal [6]. The waveband conversion function of an OPA makes it possible to use commercial EDFAs to amplify the arbitrary-band WDM signal by converting it to conventional C/L-band [7]. Moreover, by applying the waveband conversion to transceivers, the multiband signal can be transmitted/detected using only conventional C/L-band transceivers [8]. We configured the lumped multiband repeater with two hybrid OPA/EDFA configurations supporting S- and U-bands to yield high output power. Moreover, the fibre launch power and spectrum tilt were designed based on a Gaussian noise (GN) model accounting for the wideband ISRS effect [9]. Fig. 1 shows a comparison of this work with recent reports of ≥80-km-span inline-amplified multiband transmission [1–5,10–13]. We achieved a net bitrate of 133.06 Tbps at 1040 km over a 22.05-THz signal bandwidth in four bands without the DRA.

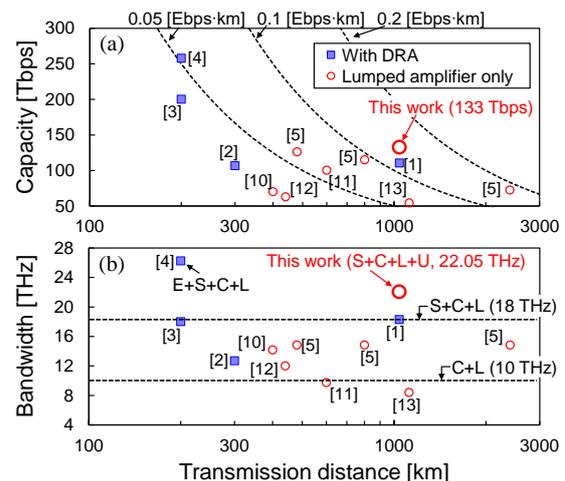

**Fig. 1:** Recent demonstrations of ≥80-km-span inline-amplified multiband transmission over SMF link. (a) Capacity or (b) inline-amplification bandwidth v.s. transmission distance.

## Experimental setup

Fig. 2 shows the experimental setup. The channel under test (CUT) was modulated with 144-Gbaud polarization-multiplexed probabilistic constellation shaped (PCS) QAM using the bandwidth-doubler-based transmitter [14]. We used

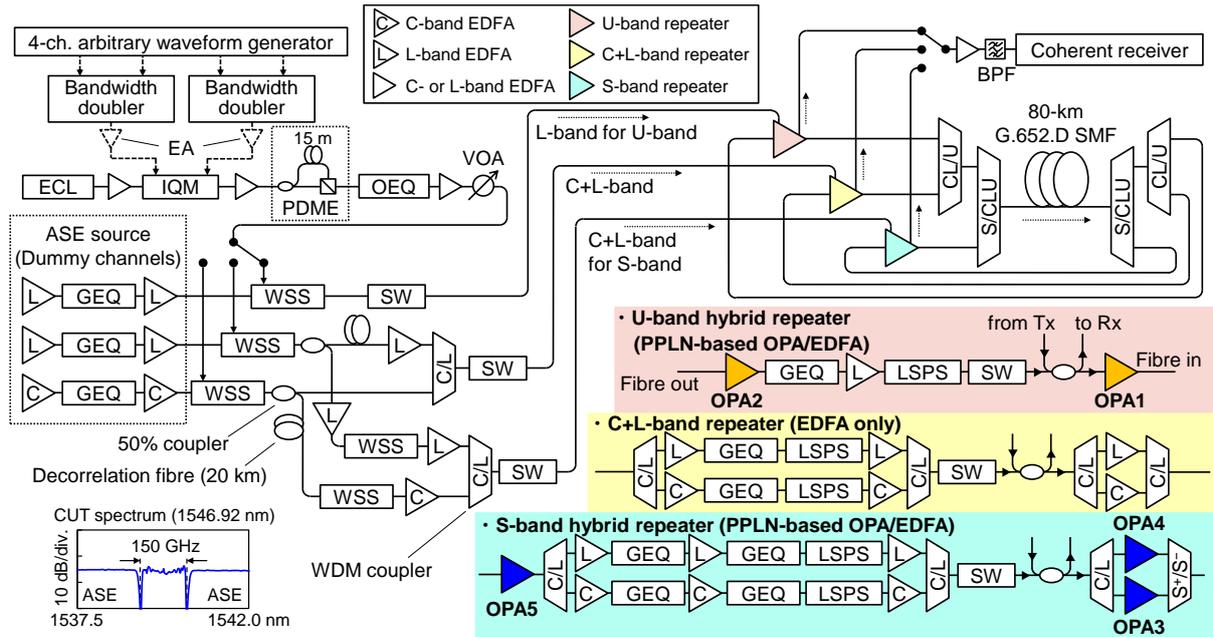

**Fig. 2:** Experimental setup with three loops for S+C+L+U-band transmission. EA: electrical amplifier, ECL: external cavity laser, IQM: I/Q modulator, PDME: polarization-division-multiplexing emulator, OEQ: optical equalizer, VOA: variable optical attenuator, SW: optical switch, BPF: bandpass filter, and LSPS: loop-synchronous polarization scrambler.

PCS-64QAM for the C/L/U-bands and PCS-16QAM for the S-band, of which entropies per polarization were 5.053 and 2.923 bits, respectively. The CUT was calibrated with digital pre-processing and optical equalization [15]. The other channels were emulated with amplified spontaneous emission (ASE) from C- and L-band EDFAs with 150-GHz channel spacing. The CUT and the ASE-based WDM channels were combined with a wavelength-selective switch (WSS). The ASEs for the C- and L-band WDM signals were shaped within 1529.75–1564.48 nm (4.35 THz) and within 1569.18–1620.06 nm (6.00 THz), respectively. The U- and S-band WDM signals were generated by converting an L-band signal and a C+L-band signal in the recirculating loop with PPLN-based OPAs, respectively. The L-band ASE for the U-band WDM signal was shaped within 1561.22–1601.67 nm (4.8 THz).

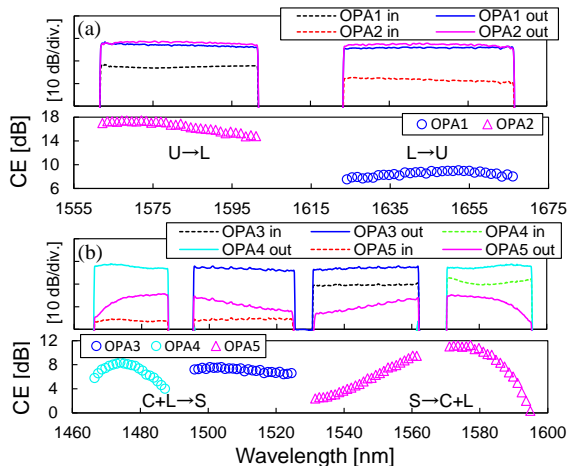

**Fig. 3:** Input/output spectra and conversion efficiency (CE) of each OPA in (a) U-band repeater and (b) S-band repeater.

The ASE for the S-band WDM signal was generated by splitting the C+L-band WDM signal. Both split signals were decorrelated with 20-km SMFs. The S-band signal was divided into two bands, one generated from the L-band signal ($S^-$) and the other generated from the C-band signal ($S^+$). The L- and C-band ASEs for the $S^-$- and $S^+$-band WDM signals were shaped by WSSs within 1570.42–1595.49 nm (3.00 THz) and within 1530.92–1562.03 nm (3.90 THz), respectively.

The transmission line consisted of three recirculating loops for the S-band, C+L-band, and U-band WDM signals. The transmission fibre was an 80-km G.652.D SMF. We used phosphorus co-doped silicate EDFAs to amplify the L-band signals [16]. Each WDM signal was input to the loop from the midpoint of each repeater and multiplexed/demultiplexed at the input/output of the transmission fibre. The C+L-band repeater was configured with C-band EDFAs and L-band EDFAs via WDM couplers. A gain equalizer (GEQ) was used in each band. For the U-band signal transmission, the L-band WDM signal was converted to the U-band within 1623.35–1666.67 nm before the transmission fibre by OPA1 with a degenerate wavelength ($\lambda_0$) of 1612.44 nm [17]. Each OPA was in a polarization-diverse configuration [6]. At the output of the fibre, the U-band WDM signal was reconverted to the L-band with OPA2. The reconverted signal was amplified and equalized by an L-band EDFA and a GEQ. Fig. 3(a) shows the input/output spectra and conversion efficiencies (CEs) of OPA1 and 2. Outputs of the OPAs included the amplified signal at the original band and the wavelength-converted idler.

OPA1 and 2 had average CEs of +8.4 dB and +16.5 dB, respectively, which were used as part of the repeater gain. The original signal components were rejected by WDM couplers. For the S-band signal transmission, OPA3, 4, and 5 with a $\lambda_0$ of 1527.99 nm [6] were used to inter-convert the WDM signal between the C+L-band and the S-band. Fig. 3(b) shows the input/output spectra and CEs of the three OPAs. To increase the launch power, two OPAs (OPA3&4) were used to convert the C+L-band WDM signal to the S-band within 1465.98–1487.80 nm and 1495.41–1525.08 nm before the transmission fibre. The CE spectra of these OPAs were made complementary by detuning the waveguide temperature [18]. After transmission, the S-band signal was simultaneously reconverted to the C+L-band with the OPA5, in which the waveguide temperature was detuned to achieve positive CE over the entire S-band signal (~8 THz from edge to edge). The positive CE of OPA5 suppressed the degradation of OSNR due to waveband conversion. The reconverted WDM signal was separated into C- and L-bands and amplified by EDFAs.

All channels were received in the C- or L-band with the coherent receiver including a C- or L-band EDFA and four 70-GHz balanced photodetectors. The received signal was digitized using a digital oscilloscope at 256 Gsample/s and demodulated offline based on a frequency domain 8×2 adaptive equalizer with a pilot-based carrier phase recovery same as in [19]. The pilot overhead was 1.59%.

**Results**

Fig. 5(a) shows the input/output spectra of the transmission fibre and launch powers of each band. The S-band WDM signal requires high launch power to compensate for the excessive attenuation due to the ISRS effect. The OPA/EDFA hybrid repeater achieved a launch power of 23.0 dBm for the 6.9-THz S-band signal. We calculated the optimum launch power and spectrum tilt on the basis of a closed-form GN model accounting for the ISRS [9], which was performed with the S-band launch power fixed at 23.0 dBm. We roughly adjusted the experimental launch power and spectrum tilt of each band to the calculated ones. The total launch power was 25.5 dBm. Fig. 5(b) shows the attenuation spectra of the transmission fibre measured by sweeping the CW laser (i.e., without ISRS) and launching the S-to-U WDM signal. The shortest-wavelength channel suffered from an excess attenuation of 5.7 dB due to the ISRS, while the longest-wavelength channel obtained a 4.5-dB gain.

Fig. 6 shows the net bitrates of all channels after 1040-km (13×80 km) transmission calculated in the same manner of rate adaptive coding as in [20] and the same forward error correction codes as in [19]. The minimum code rates for the PCS-64QAM and PCS-16QAM channels were 0.668 and 0.544, respectively. The U-band channels were comparable to the C/L-band channels because of the designed ISRS effect. The achieved total net bitrate was 133.06 Tbps with an average channel net bitrate of ~905 Gbps. The performance in the S-band could be further improved by applying the DRA to the link.

**Conclusion**

We demonstrated inline-amplified transmission over 22.05 THz within the S-to-U-band without DRA and achieved the highest net bitrate of 133.06 Tbps in >1000-km transmission with ≥80-km fibre spans. This result showed the potential to extend the signal bandwidth in terrestrial fibre-optic transmission systems to 22 THz using the C/L transceiver configuration and EDFAs with the PPLN-based OPAs.

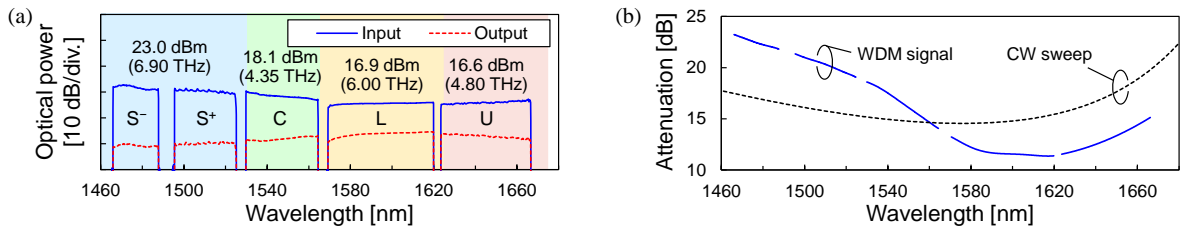

**Fig. 5:** Transmission conditions. (a) Input/output spectra of transmission fibre (80-km G.652.D SMF). (b) Attenuation spectra of transmission fibre measured by sweeping CW laser (without ISRS) and launching WDM signal (with ISRS).

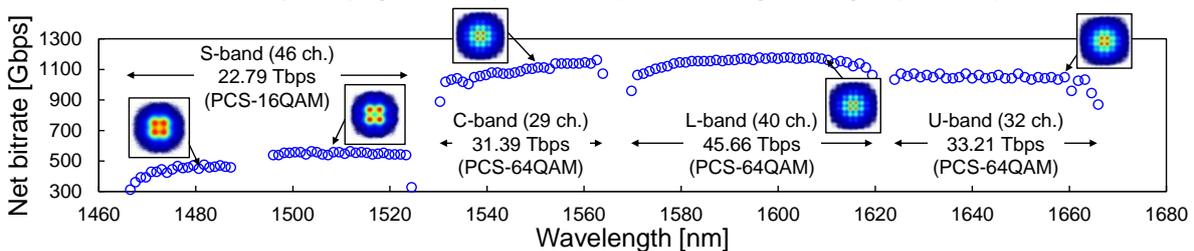

**Fig. 6:** Net bitrates of all 147 channels after 1040-km (13×80 km) transmission (total net bitrate: 133.06 Tbps).

**Acknowledgements**

Part of this work was obtained from commissioned research (JPJ012368C04501) and the grant program (JPJ012368G60101) from National Institute of Information and Communications Technology (NICT), Japan.